\documentclass[twocolumn, showkeys]{revtex4}
\usepackage{graphicx}

\begin{document}

\title{Superconductivity of Ta$_{34}$Nb$_{33}$Hf$_{8}$Zr$_{14}$Ti$_{11}$ high entropy alloy from first principles calculations}

\author{K. Jasiewicz}
\author{B. Wiendlocha}
\email[email: ]{wiendlocha@fis.agh.edu.pl}
\author{P. Korben}
\author{S. Kaprzyk}
\author{J. Tobola}

\affiliation{AGH University of Science and Technology, Faculty of Physics and Applied Computer Science, Al. Mickiewicza 30, 30-059 Krakow, Poland}

\date{\today}

\begin{abstract}
The Korringa-Kohn-Rostoker method with the coherent potental approximation (KKR-CPA) is applied to study the first superconducting high entropy alloy (HEA) Ta$_{34}$Nb$_{33}$Hf$_{8}$Zr$_{14}$Ti$_{11}$ (discovered in 2014 with $T_c$ = 7.3 K), focusing on estimations of the  electron-phonon coupling constant $\lambda$. Electronic part of $\lambda$ has been calculated 
using the rigid muffin-tin approximation (RMTA), while the phonon part has been approximated using average atomic mass and experimental Debye temperature. The estimated $\lambda$=1.16 is close to the value determined from specific heat measurements, $\lambda$=0.98, and suggests rather strong electron-phonon coupling in this material.
\keywords{High entropy alloy, disorder, superconductivity, KKR-CPA}

\end{abstract}

\maketitle

\section{Introduction}
First reports about superconducting binary solid solutions of transition elements appeared in early 1960s~\cite{b1}. Those alloys with general formula A$_{x}$B$_{1-x}$ (A, B = Hf, Zr, Ti, Ta, V, Cr, Mo, Nb, Re) have moderate transition temperatures, which do not exceed 12 K (Mo-Re alloy). Recently, it has been reported~\cite{b2} that superconductivity also appears in more complex metallic alloys, commonly called {\it high entropy alloys} (HEA) due to particular role of configuration entropy in their crystal stability~\cite{b3}.
Combination of Ta, Nb, Hf, Zr and Ti with concentrations of 0.34, 0.33, 0.08, 0.14, and 0.11~\cite{b2} respectively, resulted in discovery of II - type superconductor with body-centered cubic structure (a = 3.36~\AA). Transition temperature of Ta$_{34}$Nb$_{33}$Hf$_{8}$Zr$_{14}$Ti$_{11}$ (TNHZT) is about $T_c = 7.3$~K, upper critical field $\mu_0H_{c2}$ $\approx$ 8.2 T, lower critical field $\mu_0H_{c1}$ $\approx$ 32 mT and energy gap in the electronic density of states at Fermi level 2$\Delta \approx$ 2.2 meV $\simeq 3.5$~k$_BT_c$. 
It is worth noting, that the number of valence electrons per ''atom'' (VEC) is slightly below the half filling of the $d$-shell, being equal to 4.67, which is also consistent with the value of 4.7, for which Hulm {\it et al.}~\cite{b1} observed maximum transition temperature of the binary superconducting solid solutions $A_{x}B_{1-x}$ (second maximum was observed near 6.4). 
This new class of alloys (HEA)~\cite{b3} are frequently called a ''metallic glass on an ordered crystal lattice''~\cite{b2} since, in spite of a relatively large number of elements (from five to more than twenty) with concentrations ranging from 5 to 35\%, HEA crystallize in surprisingly simple structures. Instead of complex ordered intermetallic phases, random $bcc$ or $fcc$ structures are formed, and recently $hcp$ HEA has been reported~\cite{b5}. 

The aim of this work is to calculate the electronic structure of the superconducting TNHZT system and to estimate the electron-phonon coupling constant $\lambda$. Such calculation should allow verifying whether the superconductivity of the system can be explained using the conventional mechanism as suggested from experimental data~\cite{b2}.

\section{Method of calculation}
A high degree of chemical disorder occuring in HEA, i.e. five or more atoms randomly occupying the same crystallographic site, complicate the calculations of the electronic structure and resuling physical properties of these materials. 
Fortunately, the Korringa-Kohn-Rostoker method combined with the coherent potential approximation (KKR-CPA)~\cite{b15,b13} appears to be well adapted technique to treat such complex cases, since it allows to perform first principles computations of chemically disordered materials in a self-consistent way. The random arrangement of constituent elements (five in TNHZT), as observed in real HEA systems, is imitated by an ordered medium of ''CPA atoms'', actually representing an average over all possible configurations of the disordered lattice (without necessity to construct supercells). CPA, developed since 1960s~\cite{soven}, together with KKR showed to be highly efficient technique in describing electronic structure and related physical properties (e.g. magnetic, 
thermoelectric, superconducting) of disordered solids~\cite{ebert}.  
Interestingly, KKR-CPA has already been applied to calculate electronic structure, formation energy, phase preference and magnetic properties of selected HEA~\cite{b11,stocks-2016}. Here, this method is used to compute the electronic contribution to the electron-phonon interaction. Crystal potential of the {\it muffin-tin} form was constructed using the local density approximation (LDA), with Barth - von Hedin parametrization~\cite{lda} in the semi-relativistic approach. Angular momentum cut-off was set to $l_{\rm max}$ = 3 and highly converged results were obtained for about 450 k-points grid in the irreducible part of Brillouin zone (IRBZ) for the self-consistent cycle and 2800 IRBZ k-points for the densities of states (DOS) computations. The Fermi level ($E_F$) in disordered alloys, which is particularly important when accounting for superconducting parameters depending on DOS values, was accurately determined from the generalized Lloyd formula~\cite{b14}.

\begin{table}[t]
\caption{Electronic properties of Ta$_{34}$Nb$_{33}$Hf$_{8}$Zr$_{14}$Ti$_{11}$. $M_i$ is given in u, $N$($E_F$) in Ry$^{-1}$, $\eta$ in mRy/a${_B^2}$ ($a_B=0.529$ \AA).}
\begin{tabular}{lllllll}
\hline
 &  $M_i$ & $N$($E_F$)& $\eta_i$  &$\eta_{sp}$ &$\eta_{pd}$ &$\eta_{df}$ \\
\hline
Ta  & 181&17.2 & 154.0 & 0.9 & 50.7 & 102.4 \\
Nb  & 93 &18.8 & 161.9 & 4.7 & 55.3 & 102.0 \\
Hf & 179 &16.2 & 160.5 & 1.7 & 71.4 & 87.5\\
Zr  & 91 &17.2 & 176.1 & 7.1 & 79.3 & 89.7 \\
Ti  & 41 &26.6 & 126.2 & 4.9 & 45.7 & 75.6\\
\hline
  \end{tabular}
  \label{tab1}
\end{table}

The electron-phonon coupling (EPC) parameter $\lambda$, roughly speaking, accumulates the interaction of all conduction electrons at the Fermi surface with phonons in multi-atomic system. When applying the widely used Rigid Muffin Tin Approximation (RMTA)~\cite{rmta,rmta3}, EPC may be decoupled into a sum of individual atomic-dependent contributions,
\begin{equation}\label{eq:lambda}
\lambda = \sum_i \frac{\eta_i}{M_i\langle{\omega_i^2}\rangle},
\end{equation}
where $\eta_i$ is the $i$-th atom's McMillan-Hopfield (MH) parameter~\cite{mcm,hop} representing the electronic contribution to EPC, $M_i$ is the  atomic mass, and $\langle{\omega_i^2}\rangle$ is the appropriately defined ''average square'' atomic vibration frequency. 
For a more detailed discussion of the approximations involved in the aforementioned methodology, see e.g. Refs.~\cite{rmta3,prb-bw06}, whereas recent results obtained within this approach can be found in \cite{bw-pss06,bw-prb08,bw-intermet14,bw-prb15,papa-prb15}.

\begin{table}[t]
\caption{McMillan-Hopfield parameters (mRy/a${_B^2}$) for the elemental solids, per primitive cell (values for {\it hcp} are per 2 atoms).}
\begin{tabular}{llllll}
\hline
 & Ta ({\it bcc}) & Nb ({\it bcc}) & Hf ({\it hcp}) & Zr ({\it hcp}) & Ti ({\it hcp}) \\
\hline
$\eta$ & 155 & 152 & 54  &  62 & 48  \\ 
\hline
  \end{tabular}
  \label{tab2}
\end{table}

The McMillan-Hopfield parameters can be calculated for each atom constituting ordered crystals as well as disordered alloys, since they directly depend on the electronic structure of the system.
In general, $\eta$ is defined as $\eta = 2M\int \omega \alpha^2F(\omega) d\omega$, where $\alpha^2F(\omega)$ is the Eliashberg electron-phonon coupling function~\cite{mcm}, however since $\alpha^2F(\omega) \propto \omega^{-1}$, the dependence of $\eta$ on the phonon frequency $\omega$ cancels out, and $\eta$ becomes purely an electronic factor. In the RMTA it may be computed using the formula:~\cite{kaprzyk-nb}
\begin{equation}
\label{eq:eta} \eta_i =\!\sum_l \frac{(2l + 2)\,n_l\,
n_{l+1}}{(2l+1)(2l+3)N(E_F)} \left|\int_0^{R_{\mathsf{MT}}}\!\!r^2
R_l\frac{dV}{dr}R_{l+1} \right|^2\!,
\end{equation}
where $V(r)$ is the self-consistent potential at site $i$, $R_\mathsf{MT}$ is the radius of the $i$-th MT sphere, $R_l(r)$ is a regular solution of the radial Schr\"odinger equation (normalized to unity inside the MT sphere), $n_l(E_F)$ is the $l$--th partial DOS per spin at the Fermi level $E_F$, and $N(E_F)$ is the total DOS per primitive cell and per spin. All these quantities may be calculated for an alloy using KKR-CPA method, and each atom's contribution to EPC may be weighted by its atomic concentration $c_i$ (see, e.g.~\cite{kaprzyk-nb}).

\begin{figure}[b]%
\includegraphics[width=\linewidth]{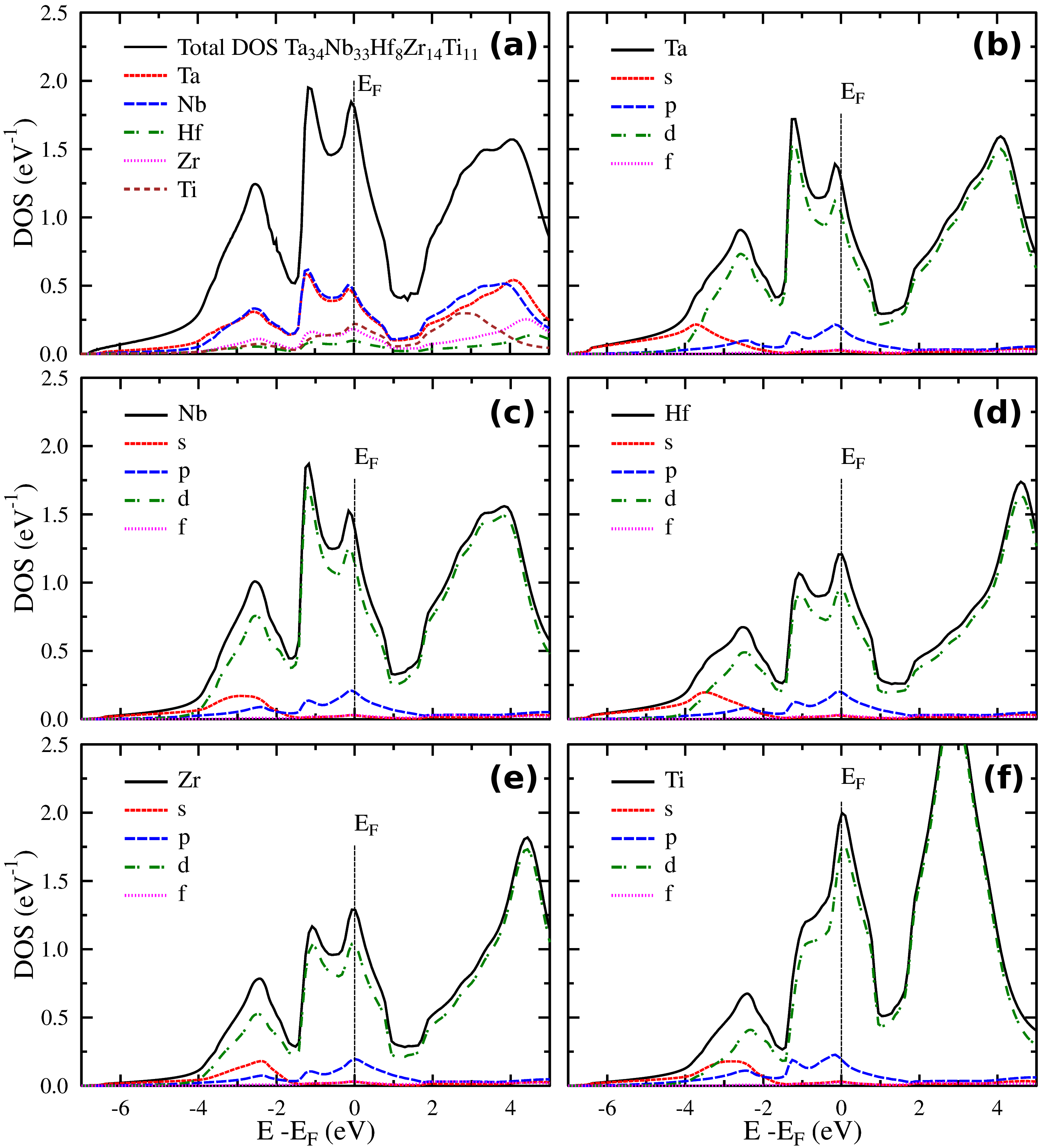}
\caption{%
Electronic density of states (DOS) of Ta$_{34}$Nb$_{33}$Hf$_{8}$Zr$_{14}$Ti$_{11}$. Panel (a): total DOS, plotted with black solid line, and partial atomic densities, marked by colors and weighted by their atomic concentrations. Panels (b)-(f): partial atomic densities with angular momentum decomposition.}
\label{fig:dos}
\end{figure}

Estimating the denominator of Eq.~(\ref{eq:lambda}), i.e. the phonon contribution to EPC, will be more difficult. 
In ordered crystals, $\langle{\omega_i^2}\rangle$ may be calculated using the computed phonon density of states~\cite{prb-bw06,bw-pss06,bw-prb08,bw-intermet14,bw-prb15}. 
For disordered alloys situation gets more complicated, since phonon computations for such a multicomponent disordered material would require studying a large number of supercells with different atomic arrangements and averaging the results over representative configurations.
Methods based on the coherent potential approximation for phonons are still under development~\cite{cpa-ph}.
Fortunately, our studied material is a simple ''monoatomic'' $bcc$ structure, which should presumably have a simple ''average'' phonon spectrum. For monoatomic crystals, $\langle{\omega^2}\rangle$ was frequently (and successfully) approximated using the experimental Debye temperature $\Theta_D$: 
$\langle{\omega^2}\rangle = \frac{1}{2}\Theta_D^2$~\cite{kaprzyk-nb,b26,b27}.
Thus, it is reasonable to assume that the denominator of Eq.~(\ref{eq:lambda}) can be approximated using concentration-average atomic mass and experimental Debye temperature: $M_i \langle{\omega_i^2}\rangle \simeq \langle{M}\rangle{1\over 2}\Theta_D^2$, where $\langle{M}\rangle = \sum_i c_iM_i$.
Setting equal $M_i \langle{\omega_i^2}\rangle$ for all the atoms in our random structure can be additionally supported by the fact, that it is proportional to the interatomic force constants (IFC). If, on average, our HEA is a homogeneous $bcc$ structure with no important structural distorsions, IFC should not differ between the sites, occupied by different atoms in the real disordered material. Thus finally, the following formula will be used for calculating the EPC constant $\lambda$ in the studied alloy:
\begin{equation}
\lambda = \frac{{\sum_i c_i \eta_i}}{{1\over 2}\langle{M}\rangle\Theta_D^2}.
\end{equation}

\section{Results and discussion}
Total electronic density of states is presented in Fig.~\ref{fig:dos}(a) along with contribution of each constituent atom. The Fermi level $E_F$ is located in the peak of total DOS, as well as partial DOSes (panels (b)-(f)), which is usually favorable for superconductivity. The values of DOS at $E_F$ are presented in Table~\ref{tab1}.
The highest contribution to the total DOS comes from Ta and Nb atoms, due to their highest atomic concentrations in the alloy. The shape of Ta and Nb DOS are similar to each other, just as the shape of Hf and Zr DOSes. The only $3d$ element, Ti, has the most pronounced DOS peak near $E_F$ and then exhibits the highest atomic $n(E_F)$. 

\begin{figure}[b]%
\includegraphics*[width=\linewidth]{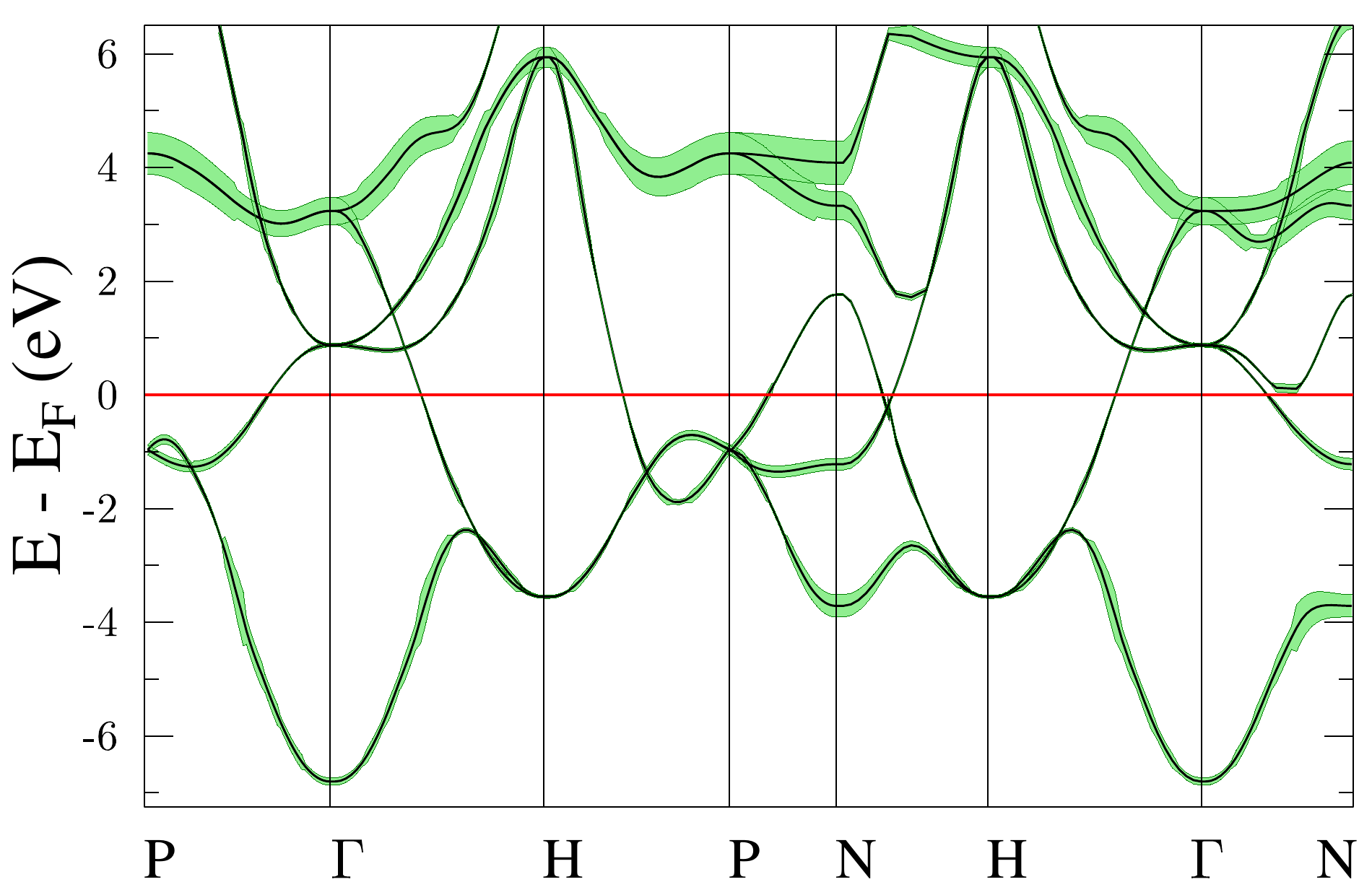}
\caption{%
Electronic dispersion relations for Ta$_{34}$Nb$_{33}$Hf$_{8}$Zr$_{14}$Ti$_{11}$. Black solid line shows the real part of the energy i.e. the band center. The bandwidth is marked in green and corresponds to the imaginary part of the energy.}
\label{fig:bnd}
\end{figure}

Fig.~\ref{fig:bnd} shows the electronic bands. Note, that in the disordered system the generalization of the usual dispersion $E({\bf k})$ relations is needed in order to take into account the ''alloy'' scattering of electrons. To describe $E({\bf k})$ in an alloy, one can use the Bloch spectral functions~\cite{ebert}, which are especially useful when strong (resonant) electron scattering takes place and well-defined bands do not exist~\cite{bw-prb13}. If the scattering is not very strong and primarily leads to the band smearing, the complex energy band technique~\cite{butler-kubo,stopa06,scripta16} may be used. The real part of such a complex energy eigenvalue shows the center of the band, whereas the imaginary part describes the band width, which corresponds to the finite life time of the electronic state $\tau = \frac{\hbar}{2{\rm Im}(E)}$.
In the studied case of the high entropy alloy, in spite of the large disorder between five elements on a single crystal site, electronic bands appeared to be quite sharp (Fig.~\ref{fig:bnd}). Especially near $E_F$, the bandwidth is really small, resulting in $\tau \simeq 0.5 - 1 \times10^{-14}$~s. Such values are of the same order as in typical transition-metal alloys and explain why the residual resistivity of TNHZT alloy ($\rho_0 = 36 \mu\Omega$ cm) is not much larger than (or close to) the $\rho_0$ values of binary alloys near 50-50 atomic concentrations (e.g. 25 in Ti-Hf, 45 in Ti-Zr, 12 in Zr-Hf, 10 in Nb-Ta~\cite{march} and 45 in Nb-Zr~\cite{jap-alloys}, [all values in $\mu\Omega$ cm]), and even smaller than in some ternary alloys (e.g. Ti-Zr-Nb alloys exhibit $\rho_0 \simeq 40-100 \mu\Omega$ cm depending on the composition~\cite{jap-alloys}).

\begin{figure*}[t]%
\includegraphics*[width=0.8\linewidth]{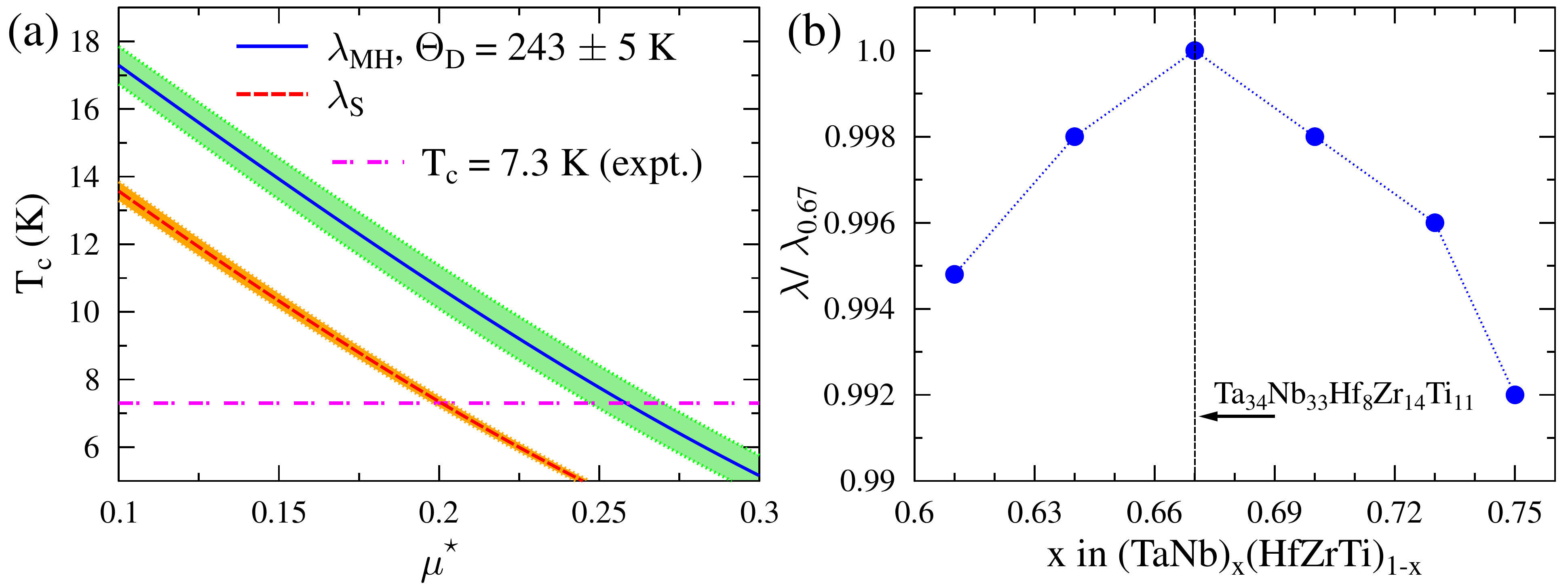}
\caption{%
(a) $T_c$ as a function of $\mu^*$, computed using the ''McMillan-Hopfield'' $\lambda_{\rm MH}$ = 1.16 $\pm$ 0.05, and ''Sommerfeld'' $\lambda_{\rm S}$ = 1.0 (see, text). (b) Change in $\lambda$ when the relative composition between group-4 (Ti,Zr,Hf) and group-5 (Nb,Ta) elements is modified, changing the VEC between 4.6 and 4.75.}
\label{fig:x}
\end{figure*}

Noteworthy, the total $N(E_F) = 24$ Ry$^{-1}$ = 1.76~eV$^{-1}$ corresponds to the ''bare'' value of the Sommerfeld electronic specific heat coefficienf $\gamma_0$ = 4.2~(mJ mol$^{-1}$K$^{-2}$), which, matched to the experimental value of $\gamma_{\rm expt}$ = 8.3 $\pm$ 0.1~(mJ mol$^{-1}$K$^{-2}$), yields the EPC parameter $\lambda$ = $\gamma_{\rm expt}/\gamma_{0}$ - 1 = 0.98 $\pm$ 0.01 $\simeq$ 1.0, suggesting relatively strong electron-phonon coupling.
Table~\ref{tab1} also shows the computed MH parameters.
Despite quite low DOS at $E_F$, Zr atoms have the largest MH parameter. Hovewer, more important when calculating electron - phonon coupling strength are $\eta$ of atoms with highest concentrations i.e. tantalum and niobum. 
For theses two elements, the most important scattering channel is $d$-$f$, typically for transition elements. In zirconium and hafnium, $p$-$d$ and $d$-$f$ scattering channels are equally relevant.

Interesting observation can be made, if the values of atomic $\eta_i$ are compared to their values computed in elemental crystals, shown in Table~\ref{tab2} (values for $hcp$ metals with two atoms per primitive cell are doubled). Ta and Nb, having the same $bcc$ crystal structure as the TNHZT alloy, have similar values of $\eta_i$,
whereas for Hf, Zr and Ti in the HEA alloy they are 2.5 - 3-times larger. This feature results from both, change of the unit cell geometry (strongly changing DOS shape) and the shift of the Fermi level position due to the change in VEC, which locates $E_F$ in the local peak of DOS. The sole change of the crystal structure from $hcp$ to $bcc$, with $a$ = 3.36~\AA, roughly doubles their $\eta_i$. 

Now, taking the experimental value of $\Theta_D = 243 \pm 5$ K of the alloy~\cite{b2}, average atomic mass of $\langle M \rangle = 124.55$  a.u. and MH parameters $\eta_i$ from Table~\ref{tab1}, one arrives at the EPC parameter $\lambda = 1.16 \pm 0.05$.
This result is in a good agreement with the value estimated above using the Sommerfeld coefficient, $\lambda = 1.0$.
$T_c$ may be now calculated using the McMillan formula~\cite{mcm}
\begin{equation}
T_c = \frac{\Theta_D}{1.45}\exp\left[-\frac{1.04(1+\lambda)}{\lambda-\mu^*(1+0.62\lambda)}\right].
\end{equation}
Taking the standard value of the Coulomb pseudopential, $\mu^*$ = 0.13, calculated critical temperature is $T_c$ = 15 K, when $\lambda = 1.16$ is used, two times higher than the experimental $T_c = 7.3$~K. Fig.~\ref{fig:x} shows $T_c$ plotted as a function of $\mu^*$, with shading area corresponding to the uncertainty in $\Theta_D = 243 \pm 5$~K, used for both $\lambda$ and $T_c$ computation. As can be seen, the value of $\mu^*$ as large as 0.25 has to be used to obtain $T_c$ consistent with experimental data. 
This discrepancy may be caused by the rough estimation of the phonon part of the EPC constant. 
Slightly smaller difference between theory and experiment is observed, if $\lambda$ = 1.0, obtained from the analysis of the Sommerfeld coefficient $\gamma$, is used ($T_c$ = 11.5~K), however it also requires applying rather 
large $\mu^*$ = 0.20 to reach the experimental $T_c$ (see, Fig.~\ref{fig:x}).  
It is worth reminding here that similar problems with large $\mu^*$ needed to reproduce experimental $T_c$ were reported in literature for various materials, like Nb$_3$Ge ($\mu^*$=0.24)~\cite{b28}, V ($\mu^*$=0.3)~\cite{b29} or MgCNi$_3$ ($\mu^*$=0.29)~\cite{durajski}.
Nevertheless, presented KKR-CPA calculations, within RMTA approach, support the electron-phonon coupling mechanism of the superconductivity in the TNHZT HEA, and show relatively strong electron-phonon interaction, with $\lambda \simeq 1$.

Finally, to investigate how the superconductivity in the studied system is sensitive to moderate changes in composition, we have calculated variations in MH parameters when VEC is modified from 4.67, in the range 4.6-4.75, by changing the relative amount of group-4 (Ti,Zr,Hf) and group-5 (Nb,Ta) elements, and keeping unchanged the ratio between elements in each group. Assuming, that the IFCs does not change (i.e. assuming $\langle M \rangle \Theta_D^2$ = const.) the change in $\lambda$ was simulated and is plotted in Fig.~\ref{fig:x}(b). First of all, $\lambda$ is not very sensitive to changes in composition in this range, but what is quite intriguing, it is maximized for this particular ratio of group-4 and group-5 elements, investigated experimentally~\cite{b2}.

\section{Summary}
KKR-CPA electronic structure calculations of the superconducting high entropy alloy Ta$_{34}$Nb$_{33}$Hf$_{8}$Zr$_{14}$Ti$_{11}$ accounting for chemical disorder as random distribution of constituent atoms in $bcc$ unit cell, are reported. The electron-phonon coupling constant $\lambda$ = 1.16 was calculated, using the computed McMillan-Hopfield parameters and experimental Debye temperature. The obtained $\lambda$ well corroborates with the value of $\lambda \sim$ 1.0 extracted from the experimental electronic specific heat and calculated electronic DOS at $E_F$.
In view of our results the TNHZT alloy can be classsified as the strong electron-phonon coupling superconductor.

\section{Acknowledgements}
This work was partly supported by the Polish Ministry of Science and 
Higher Education. J.T. and B.W. also acknowledge support from the 
Accelerated Metallurgy FP7 Project (contract NMP4-LA-2011-263206) 
coordinated by the European Space Agency and by the individual partner 
organizations.

%
%

\end{document}